 \documentclass[preprint2]{aastex}
 \usepackage{graphicx}

\shorttitle{An ancient merger in M31?}
\shortauthors{Hammer et al.}

\begin{document}


\title{Does M31 result from an ancient major merger? }


\author{F. Hammer\altaffilmark{1}, Y. B. Yang\altaffilmark{2}, J. L. Wang\altaffilmark{1,2}, M. Puech\altaffilmark{1}, H. Flores\altaffilmark{1} \& S. Fouquet\altaffilmark{1}}
\affil{Laboratoire GEPI, Observatoire de Paris, CNRS-UMR8111, Univ. Paris-Diderot 5 place Jules Janssen, 92195 Meudon France}
\email{francois.hammer@obspm.fr}


\altaffiltext{1}{Laboratoire GEPI, Observatoire de Paris, CNRS-UMR8111, Univ. Paris-Diderot, 5 place Jules Janssen, 92195 Meudon France}
\altaffiltext{2}{NAOC, Chinese Academy of Sciences, A20 Datun Road, 100012 Beijing, China}

%Received 2010 August 2; accepted 2010 September 29


\begin{abstract}
The M31 haunted halo is likely associated with a rich merger history, currently assumed to be caused by multiple minor mergers. Here we use the GADGET2 simulation code to test whether M31 could have experienced a major merger in its past history. Our results indicate that a (3$\pm$0.5):1 gaseous rich merger with $r_{pericenter}$=25$\pm$5 kpc and a polar orbit can explain  many properties of M31 and of its halo. The interaction and fusion may have begun 8.75$\pm$0.35 and 5.5 $\pm$0.5 Gyr ago, respectively. Observed fractions of bulge, and the thin and thick disks can be retrieved for a star formation history that is almost quiescent before the fusion. This also accords well the observed relative fractions of intermediate age and old stars in both the thick disk and the Giant Stream. In this model, the Giant Stream is caused by returning stars from a tidal tail which contains material previously stripped from the satellite prior to the fusion. These returning stars are trapped into elliptical orbits or loops for long periods of time which can reach a Hubble time, and are belonging to a plane that is 45 degrees offset from the M31 disk position angle. Because these streams of stars are permanently fed by new infalling stars with high energy from the tidal tail, we predict large loops which scale rather well with the features recently discovered in the M31 outskirts.

We demonstrate that a single merger could explain first-order (intensity and size), morphological and kinematical properties of the disk, thick disk, bulge and streams in the halo of M31, as well as the observed distribution of stellar ages, and perhaps metallicities. This challenges the current scenarios assuming that each feature in the disk (the 10 kpc ring) or in its outskirts (thick disk, the Giant Stream, and the numerous streams) is associated with an equivalent number of minor mergers. Given the large number of parameters, further constraints are certainly required to better render the complexity of M31 and of the substructures within its halo which may ultimately lead to a more precise geometry of the encounter. This would allow us, in principle, to evaluate the impact of such a major event on the Andromeda system and Local Group.

%Our solution for which all stellar streams in the M31 halo are linked to a single event - returning of stars from a tidal tail- simply solve the issue of finding missing satellites associated to them. If true, many properties of the Andromeda system and of the Local Group may have to be revisited accordingly.

%It may also lead to a dramatic revision of the properties of the Local Group, since a fraction of its population of dwarves  may have been generated by a giant collision occurring in the past Andromeda history.
\end{abstract}





\section{Introduction}
 Our nearest giant neighbour, M31 has attracted a considerable amount of interest since the discovery of many large-scale structures that surround its outskirts. These prominent structures include the Giant Stream \citep[infall of 1.5 $\times$ $10^{8}$ M$_{\odot}$ stellar mass,][]{ibata01} and the gigantic thick disk containing about 10\% of the disk luminosity \citep{ibata05}.  The halo of M31 is haunted  by not less than 16 substructures that led \citet{tanaka10} to conjecture that they are due to as many accretion events involving dwarf satellites with mass 10$^{7}$-10$^{9}$ M$_{\odot}$ since z $\sim$ 1. Observations at radio wavelengths (\citet{westmeier05, thilker04};  and references therein) reveal high velocity clouds, whose location and kinematics partly follow the Giant Stream. Moreover, a 10kpc, pseudo-ring inserted in the M31 disk dominates star formation \citep{baade64}, HI gas \citep{roberts66}, molecular gas \citep{nieten06}, and dust emission \citep{gordon06}.
 
 Simulations of the numerous structures in the M31 outskirts have always assumed them to be caused by minor satellites. M32 is understood to be the perturber of the spiral arms \citep{byrd78, byrd83}, while NGC 205 has been modelled as the cause of the warp seen in the optical and HI disks of M31 \citep{sato86}. The thick disk has been modelled by earlier disruption, several Gyr ago, of dwarf galaxies on prograde orbits that are coplanar with the disk \citep{penarrubia06}. Simulations of the Giant Stream assumed a  collision with an unknown satellite $\sim$ 0.7 Gyr ago  \citep{font06, fardal08}. The 10 kpc ring could also be reproduced by an interaction with M32, assuming a polar orbit \citep{block06}.

 In spite of their success in reproducing M31 structures, several of these simulations may be  speculative because the M31 satellite orbits are currently unknown \citep[e.g.,][]{fardal09}. More problematic is the fact that stars in the Giant Stream \citep{brown07} have ages from 5.5 to 13 Gyr, which is difficult to reconcile with a recent collision \citep[e.g.,][]{font08}. If the substructures are formed from different progenitors, why do they show obvious similarities in metallicity \citep[e.g.,][]{ferguson05} ? The main motivation for  only investigating minor mergers is to preserve the M31 disk \citep{mori08}. But the M31 disk is not necessarily very old and permanent, as its associated stellar clusters display young to moderate ages \citep[$<$ 5-7 Gyr,][]{beasley04}, and most of the stars in the outskirt structures are older than this. One should not discard the possibility that the gigantic structures may have been formed at the same time or even earlier than the disk. In fact  many co-workers in the field \citep{rich04, ibata04, brown06} have hypothesised a possible major merger in the past history of M31\footnote{Quoting \citet{vandenBergh05}: ``Both the high metallicity of the M31 halo, and the $r^{1/4}$ luminosity profile of the Andromeda galaxy, suggest that this object might have formed from the early merger and subsequent violent relaxation, of two (or more) relatively massive metal-rich ancestral objects.''}. More recently, the kinematics of the M31 globular system has been attributed to an ancient major merger \citep[e.g.,][]{bekki10}.

Understanding the nature of M31 certainly has an impact in cosmology
because it and the Milky Way, are the only two giant
spirals in our immediate neighbourhood. In contrast to the Milky Way,
M31 has properties (absolute K luminosity, angular momentum, [Fe/H] in
the outskirts) similar to the average of large spirals having the same
rotation velocity \citep{hammer07}. Half of the progenitors of
spiral galaxies in this range were not relaxed 6 billion years ago,
according to the detailed studies of their morphologies \citep{hammer05, neichel08, delgado10} and kinematics \citep{puech08, yang08}. \citet{hammer09} verified that
various merger phases can reproduce quite well the observed
morphologies and kinematics of these non-relaxed galaxies. This agrees
with the disk rebuilding scenario \citep{hammer05}, according to which most
spiral disks have been rebuilt following a major, gas-rich merger
during the past 8-9 Gyr. High gas fractions have been shown to
be essential during this process \citep{hopkins08, hopkins10}.
Understanding the different substructures of M31 as resulting from
a single event such as a gas-rich major merger is an important step
in validating or dismissing such a channel for spiral disk formation.

The goal of this study is to investigate whether a past major merger, instead of multiple minor mergers, can reproduce most of the peculiarities of M31. The $\Lambda$CDM cosmology ($H_0$=70 km s$^{-1}$ Mpc$^{-1}$, $\Omega_M=0.3$ and $\Omega_\Lambda=0.7$) is adopted throughout the paper.

\begin{figure*}[!ht]
  \centering
  \includegraphics[width=15cm]{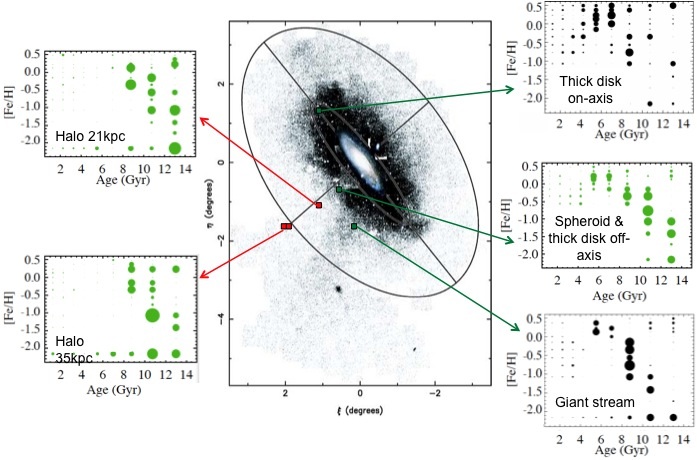}
  \caption{Chronological sketch of the structures surrounding M31. In the central panel \citep[reproduced from][]{ibata05}, the large and thick rotating disk  is a vast flattened structure with a major axis of about 4 degrees.  Squares represent fields observed by \citet{brown06, brown07, brown08}, and are linked to their measurements by arrows.}
  \label{fig1}
\end{figure*}

 \section{Observational constraints and initial conditions}

 We assume in the following that M31 results from a past major merger. Our goal is to simultaneously reproduce: (1) the M31 disk at its rotation velocity, and the bulge, with B/T=0.28; (2) the presence of a very extended, co-rotating thick disk; (3) the 10kpc dust, HI, star-forming ring; (4) the Giant Stream, and (5) the observed age distributions of stars in the different substructures. Such a choice may appear somewhat subjective although it has been adopted after thoroughly reviewing many of the M31 properties.  First, because the M31 system is very complex to model, we need to limit the number of features that can be reproduced. Second, the limitations of our model (number of particles) prevent us from describing very small structures, for example the double nucleus of M31 \citep[but see][]{hopkinsQ10}. Third we aim at reproducing the zero and first-order features of each substructure of M31, including their stellar age, mass, and kinematics (if they are available from observations), but not the detailed morphology of each of them (see discussion, section 5).

According to \citet{hopkins10} (see their Figure \ref{fig7}), a mass ratio from 0.3 to 0.5 is required to reform a new bulge with B/T$\sim$0.3 in the remnant galaxy for gas fractions\footnote{The gas fraction in the progenitors should reach a higher value than in \citet{hopkins10}, since they define the gas fraction as its value just before the fusion.} ranging between 0\% and 50\%. However, reforming a significant thin disk requires that enough gas be preserved after the fusion of the cores: it is mostly made of stars from the gaseous thin disk that immediately reforms after the collision \citep{barnes02, abadi03, governato07, hopkins09}. Thus reforming an Sb galaxy like M31 probably requires a gas fraction in excess of 50\% in the progenitors. The total baryonic mass of M31 is 1.1$\times$10$^{11}$ M$_{\odot}$ \citep{hammer07}, and at first approximation we can assume that this is the value for the sum of the two progenitor masses. In order to prevent a too
  violent relaxation at the center of the main progenitor, we adopt a
  prograde-retrograde orientation of the spin axes, which has been
  shown to be more favorable in rebuilding a thin disk after a major
  merger \citep{hopkins08}.
  
  \begin{table*}[!ht]
    \caption{Initial and Adopted Conditions for a Major Merger Model for M31.}
    \begin{center}
      \begin{tabular}{ccllll}
        \tableline\tableline
        Ingredient & Rested Rrange & Comments  & Adopted Range \\
        \tableline
        total mass & 5.5$\times$$10^{11}M_{\odot}$ & 20\% of baryons & -\\
        mass ratio & 2-4 & to reform B/T$\sim$0.3 &  2.5-3.5\\
        $f_{gas}$ Gal1 & 0.4-0.6 &  expected at z=1.5\tablenotemark{a}  & 0.6\\
        $f_{gas}$ Gal2 & 0.6-0.8 &  expected at z=1.5 &  0.8\\
        Orbit & near polar & to form the ring &  - \\
        Gal1 incy\tablenotemark{b} & 10 to 90¡ & Giant Stream  & 35-75¡\\
        Gal2 incy\tablenotemark{b} & -30 to -110¡ & Giant Stream  & -55 to -110¡\\
        Gal1 incz\tablenotemark{c} & 90 to 110¡ & Giant Stream & 90-110¡\\
        Gal2 incz\tablenotemark{c} & 90 to 110¡ & Giant Stream  & 90-110¡\\
        Spin Gal1 & prograde & - & -\\
        Spin Gal2 & retrograde & significant remnant disk\tablenotemark{d}  & - \\
        $r_{pericenter}$ & 20-30 kpc & see the text &  22-30 kpc\\
        Feedback &high-medium\tablenotemark{e} & to preserve gas  & high/varying\tablenotemark{e}\\
        $c_{star}$ & 0.004\tablenotemark{f}-0.03  & to preserve gas & 0.03\\
        \tableline
      \end{tabular}
      %% Any table notes must follow the \end{tabular} command.
      \tablenotetext{a}{\citet{daddi10} found $f_{gas}$=0.5-0.65 in galaxies with $M_{baryon}$=0.8-2.2 $10^{11}M_{\odot}$ at z=1.5}
      \tablenotetext{b}{Orientation of the angular momentum of Gal1 relative to the orbital angular momentum, y axis}
      \tablenotetext{c}{Orientation of the angular momentum of Gal1 relative to the orbital angular momentum, z axis}
      \tablenotetext{d}{Following \citet{hopkins08}}
      \tablenotetext{e}{In some simulations, the feedback is assumed to be high before fusion and later on, assumed to drop to the medium or low feedback values of \citet{cox08}; see also section 4.2)}
      \tablenotetext{f}{Another way to preserve the gas before fusion in order to allow a significant amount of gas in the disk}
    \end{center}
    \label{tbl-1}
  \end{table*}

The presence of a gaseous ring favours a polar orbit. As shown by \citet{mori08}, the Giant Stream may result from the returning
material of the tidal tail formed just before the last passage of the
satellite. A similar although longer-lived tidal tail is predicted for
a 3:1 major merger with a polar orbit, which is associated with the
passage of the secondary galaxy just before fusion. During the first
passage and until the fusion, star formation is especially enhanced
in the secondary galaxy that is harassed by the main one \citep[e.g.,][]{cox08}. If the Giant Stream was associated with returning particles
from the tidal tail formed during the second passage (near fusion), it
would contain mostly stars older than the epoch of the fusion: this is
because the star formation cannot hold for a long time within tidal
tails due to their expected dilution.

We thus propose from Fig. \ref{fig1} a chronological history of the different structures in M31, as this figure can be used as a clock for determining the occurrence of merger phases. The star formation history of the whole system is enhanced during the first passage until the fusion and then at the fusion itself \citep[see][]{cox08}. During a merger event, most of the gas and stars in the remnant outskirts have been deposited by tidal tails formed during the first passage and later, during the fusion of the cores.  A few hundred billion years after its formation, the tidal tail dilutes, provoking a natural quenching of the residual star formation \citep[see][]{wetzstein07}. Thus the age of the material brought by tidal tails provides, with a relatively small delay, the date of both the first passage and fusion times. In the following, we assume that the first passage occurred from 8.5 to 9 Gyr ago, and that the corresponding tidal tails are responsible for the halo enrichment seen in the 21 kpc and 35 kpc fields, without significant star formation more recent than 8.5 Gyr.  The thick disk has a star formation history comparable to that of the Giant Stream, and is also generated by material returning to the galaxy mostly from tidal tails generated at the fusion. Because their youngest significant population of stars has ages of 5.5 Gyr, the delay between the first passage and fusion ranges between 3 and 3.5 Gyr. This could be accommodated for by relatively large impact parameters (20 to 30 kpc).

%The last significant star formation event in each M31 substructure can be taken as a corresponding high star formation phase during the merger. The enriched halo of M31 at 35 kpc contains almost no stars younger than 8.5 Gyr and thus it may have been enriched by tidal tails at the first passage.  Because star formation within the tidal tail naturally ceased during the dilution of the structure, it naturally explain the absence of younger stars in the halo fields. Both the giant thick disk and the Giant Stream have had their last significant star formation episodes 5.5 Gyr ago, and they could originate from more recent tidal tails originated during the fusion. It implies a delay between the two events of approximately 3 Gyr, that could be accommodated for by relatively large impact parameters (20 to 30 kpc).   
 
From the above, we can settle the initial conditions for a major merger assumed to be at the origin of M31.  Table \ref{tbl-1} describes the adopted physical parameters of such an interaction. Given the huge amount of observational data, it is beyond the scope of this study to reproduce the details of all the numerous structures in the M31 system. Instead, our aim in using hydrodynamic simulations is to determine whether or not these numerous substructures can be attributed to a single major merger in the past history of M31. 

\begin{figure*}[p]
  \centering
  \includegraphics[width=15cm]{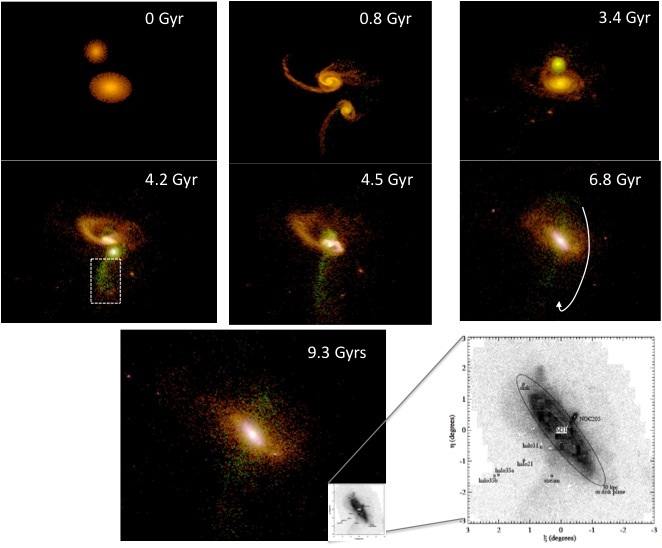}
  \caption{Different phases of a 3:1 major merger for M31 ($r_{pericenter}$=24 kpc, Gal1 incy=70, Gal2 incy=-110). The simulation starts 9.3 Gyr ago (T=0, z=1.5) and the first passage  occurs 0.7 Gyr later. Then new stars are forming (green color), especially in the secondary galaxy, until the fusion, which occurs at 4.5 Gyr. The elapsed time between the first passage and fusion is 3.8 Gyr, as the pericenter radius is large. During the second passage (T=4.2 Gyr) a tidal tail containing many newly formed, intermediate-age stars (green dots) is formed. Later on this material returns to the galaxy forming the Giant Stream, enriched with stars formed from 5 to 8 Gyr ago. The resulting galaxy in the last panel (T=9.3 Gyr) is compared with the inserted M31 image at the same scale \cite[from][, see an enlarged view of this insert on bottom-right]{ibata05}.  The Figure also illustrates the formation of the Giant Stream (see section 4.4) in a case for which the two galaxies have angle difference near the resonance (180 degrees) providing many particles stripped from the satellite at 4.2 Gyr. Some of the particles within the tidal tail have velocity below the escape velocity of the remnant system and are gradually falling back to the galaxy. They are tracing loops around the newly formed disk as indicated by the arrow in the panel at T=6.8 Gyr. Loops are fed by newly-coming particles falling back from the tidal tail and are persistent until 9.3 Gyr and later. The dotted rectangle in the 4.2 Gyr panel illustrates how we have selected the tidal tail particles (see section 4.4).}
  \label{fig2}
\end{figure*}

\section{Simulations}

We used the GADGET2 hydrodynamical code \citep{springel05}
and initial conditions similar to that of \citet{cox06, cox08}.
For the dark matter, we adopted a core density profile as in \citet{barnes02}, with a core size of 5.3 kpc and 3.06 kpc for
the main galaxy and the satellite, respectively. We verified that our results are not significantly affected by changing the density profile to a Hernquist model that is quite similar in central regions to a Navarro-Frenk-White model (see section 5). Concerning the
implementation of feedback and cooling, we have verified step by step
our ability to reproduce both the isolated and merger cases for all
the different combinations of feedback and cooling in \citet{cox08}, and as such, all parameters used in our simulations are very
similar to those of Cox et al. \citep[see][in preparation]{wang10}
Usually, the free parameters describing the star formation
  efficiency and the feedback strength are chosen in order to match the
  Schmidt-Kennicutt law between star formation and gas surface
  densities \citep[see][]{cox06}. Given the relatively large scatter
  of this relation, a large number of combinations can be accommodated
  \citep{cox06, cox08}. In the following, we explored several
  combinations that are in agreement with this relation, but with the
  additional constrain of preserving a significant gas reservoir in
  the progenitors, before fusion (see Sect. 4.2).
Properties of the progenitors are listed in Table \ref{tbl-1}, and these
galaxies have been generated to follow the baryonic Tully Fisher
relation \citep[see][ for an argument in favour of a non-evolving relation]{puech10}.
 Their atomic gas content has been assumed to be three
times more extended than the stars, as also adopted by \citet{cox06};
see also an observational support from \citet{vanderkruit07} and references therein). We
have to preserve enough gas immediately after the fusion to preserve
the formation of a significant thin disk with more than 50\% of the
baryonic mass.

During our investigations to optimise the orbital
parameters, we run the GADGET2 code with 150000 particles.
 In most simulations, we have ensured that the mass of dark matter particles
is no more than twice that of initial baryon particles, i.e., gas or
stars in the progenitors. During star formation events controlled by the gas volume
density (and following the Kennicutt-Schmidt law), we have also limited the number of newly formed stars
per gas particle to three.
This is an important limitation to the
simulation as we have verified that when newly born
stars are relatively too small in mass, they are artificially scattered due to their encountering heavy
dark matter particles. Similarly, we need to keep the mass of dark matter particles low enough to avoid non-physical disruption of the
newly reformed disk at the end of the simulation. In the following we
have tested that the decomposition of the newly formed galaxy does not
depend on the adopted number of particles within the range of 150 000-800 000. 

However, simulating the first order properties of faint structures like the Giant Stream requires a high particle number: the Giant Stream is made of 3$\times$$10^{8}M_{\odot}$, which corresponds to 1/366th of the total baryonic mass of the galaxy, and lets only a few hundreds of particles in the Giant Stream for a 150 000 particles simulation. Having fixed the range of parameters to reproduce the observed decomposition of M31 in the bulge, and the thin and thick disks, we used a limited number of simulations with 300 000 to 800 000 particles to better identify the formation of faint structures.

Table \ref{tbl-1} lists the adopted range of parameters which have been investigated (2nd column) and adopted (column 4) for several models considered in this study. These models are all part of the same family of orbital parameters, and they differ essentially by various adjustments of the star formation history or small variations of orbital parameters. We have begun with a model with moderate gas fractions (40\% and 60\% in Gal1 and Gal2, respectively) to verify whether a significant thin disk can be formed after fusion, assuming somewhat extreme conditions (e.g., high feedback and very low star formation efficiency) to prevent star formation before the fusion epoch. We then realised that most of the gas preserved before fusion forms a very thin disk \citep[see also][]{abadi03}, which also hosts some of the pre-existing stars before fusion. However, the thin disk fraction is generally not much larger than the available gas fraction before fusion. Because about 65\% of M31 stars lie in the thin disk, we have generated for our current models a higher gas fraction that allows us to test less extreme conditions for feedback and star formation efficiency. Indeed, such gas fractions are quite common at z=1.4 \citep[see][]{daddi10}, a redshift which corresponds to 9 Gyr ago, just before the interaction between the M31 progenitors (see section 2). Finally the models adopted in Table \ref{tbl-1} (column 4) are developed to optimise the reproduction of  the decomposition of the galaxy in the bulge, and the thin and thick disks, and the disk scale length (see sections 4.1 and 4.2), the presence of the 10 kpc ring (see section 4.3), the Giant Stream and its kinematics (see section 4.4), as well as the fraction of stellar ages in most of these substructures (see sections 4.2 and 4.4). An overview of the goodness of each model is provided in section 5.

Figure \ref{fig2} shows the different steps of the merger for one of our models. At the first passage (at T=0.7 Gyr), tidal tails are formed and enrich the halo, although they are mostly diluted at the fusion epoch.  Part of the material ejected during the second passage and the fusion progressively return to the galaxy after the fusion. The material associated with the tidal tail pointing to the bottom of the galaxy (see panel at T=4.2 Gyr) possesses an angular momentum very different from that of the newly formed disk. This perturbation may explain such faint features like the Giant Stream, and could be persistent for several Gyr after fusion (see section 4.4).

\section{Results and comparison with observations}

\subsection{Decomposition of the newly formed galaxy in sub-components}

\begin{figure}[h]
  \centering
  \includegraphics[width=8cm]{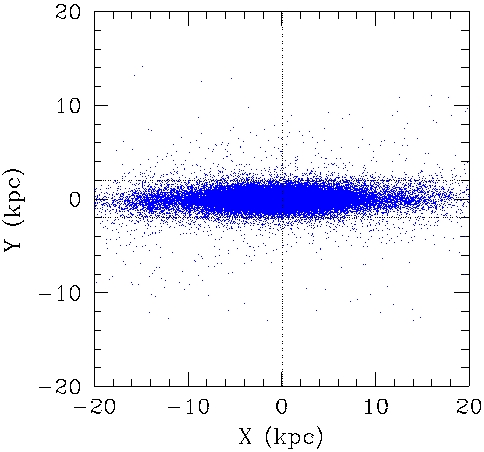}
  \caption{Distribution of newly formed stars 0.5 Gyr after the fusion which occurs 5 Gyr after the beginning of the simulation, for a 2.8:1 merger with $r_{per}$=30 kpc, Gal1 incy=65 and Gal2 incy=-90. Ten gigayears after the simulation, the diameter of the newly formed disk reaches $\sim$ 36 kpc for a thickness of 4 kpc (see thin lines) in a model for which high feedback has been assumed during the entire duration of the merger. Here the thin disk has been projected along the X-axis, and its angular momentum is oriented along the Y-axis.}
  \label{fig3}
\end{figure}

We choose to decompose the newly formed galaxy in three components,
according to the relative strength of the angular momentum in the
direction of that of the thin disk, which is approximately the orbital
angular momentum. A very accurate determination of the thin disk
angular momentum is possible because most of the preserved gas particles before the fusion naturally form
a thin disk \citep[see also][]{abadi03} in which most of the star formation occurs after fusion. We have used the
stars born 0.5 Gyr after the fusion to determine the angular momentum
of the newly formed disk (see Fig. \ref{fig3}). With this technique, we obtain an accuracy of $\pm$1
degree for determining both the disk PA and inclination.

\begin{figure}[h]
  \centering
  \includegraphics[width=8cm]{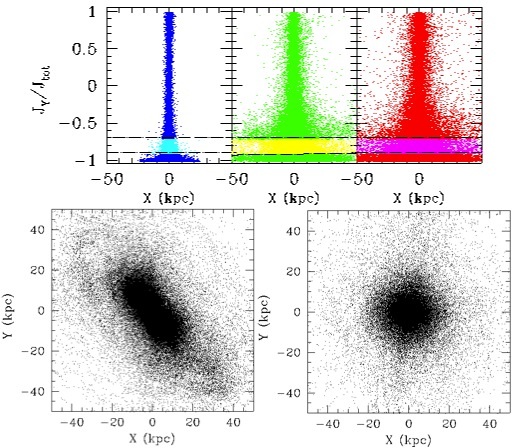}
  \caption{{\it Top:} adopted decomposition of the newly formed galaxy (same model as in Fig. \ref{fig3}), 10 Gyr after the simulation. Here the thin disk has been projected along the X-axis as in Fig. \ref{fig3}. Blue, green and red dots shown in the three panels represent stars formed 0.5 Gyr after the fusion, stars formed between the first passage (0.5 Gyr after the beginning of the simulation) to 0.5 Gyr after the fusion, and stars within the progenitors, respectively. The two dot-dashed lines delineate the thin and the thick disks, from the bottom to the top.{\it Bottom:} particles distribution in the thick disk (left panel) and in the bulge (right panel). Here the thick disk and the bulge have been projected at the right orientation of M31 thin disk (e.g., PA=38 degree and inclination=77 degrees).}
  \label{fig4}
\end{figure}

Fig. \ref{fig4} (top panels) shows the ratio of the angular momentum along the y axis  to the total angular momentum ($J_Y$/$J_{total}$) as a function of the x-axis along which the thin disk is projected (see Fig. \ref{fig3}). The final disk rotates anti-clockwise \citep[see][]{chemin09} and the thin disk (illustrated in Fig. \ref{fig3}) is easily recognisable in the Fig. \ref{fig4} top-left panel for values ranging from -0.9 and -1, which corresponds to the gaseous thin disk.  Other recently formed stars are located within a compact structure without preferred angular momentum orientations, which corresponds to the bulge.  \citet{ibata07} pointed out the confusion in the literature between the spheroid component and the thick disk. They revealed that the minor-axis region between projected radii of 7 kpc $<$ R $<$ 18 kpc is strongly affected by a rotationally-supported thick disk that corresponds to approximately one tenth of the thin disk stellar mass\footnote{It is important to notice that this region (quoted as "spheroid and thick disk off-axis" in Fig. \ref{fig1}) could well be dominated by the spheroid (Tom Brown, private communication), depending on the precise density profiles of both spheroid and thick disk.}. For a better comparison with observational data, we assume that the thick disk is made by all particles (except those of the thin disk) whose angular momentum is dominated by rotational motions along the thin disk. It means that $J_Y$ has to be larger than the combination of other components of the angular momentum (i.e., $\sqrt{J_X^2+J_Z^2}$), i.e., $J_Y$/$J_{total}$ should be larger than -1/$\sqrt{2}$ =-0.707, as illustrated by the dot-dashed lines in Fig. \ref{fig4}. We have generated the decomposition proposed by \citet{abadi03} for comparison. In their scheme, the  thick disk is the residue after subtracting the bulge and the thin disk assuming a bulge symmetrically distributed around  $J_Y$=0.  While the thin disk fraction is found to be very similar with both approaches, the residual thick disk from Abadi et al's method is almost twice that provided by our method. We notice that a significant fraction (up to 50\%) of such a residual thick disk is made of particles which are not dominated by rotation around the Y-axis, which well explains the discrepancy. For comparison with the \citet{ibata05} observations of a rotationally-supported thick disk, we keep our decomposition as described above.

\begin{table}[h]
  \begin{center}
    \caption{Decomposition in mass (unit=$10^{10}M_{\odot}$) of the newly formed galaxy 9 Gyr after the simulation, for various parameters of our modelling. For all models but the last one, high feedback has been maintained during all the simulation. For the last simulation, feedback has been abruptly dropped from high to low values \citep[from][]{cox08}, 3.5 Gyr after the beginning of the simulation and $\sim$ 0.7 Gyr after the fusion.}
    \label{tbl-2}
    \begin{tabular}{cclll}
      \tableline\tableline
      Parameters & Comp. & Thin  & Bulge & Thick \\
      & & disk & & disk\\
      \tableline
      $r_{pericenter}$=24 & Stars & 2.66 & 2.51 & 0.85 \\
      mass ratio =3.5 & Gas & 3.24 & 0.15 & 0.09\\
      Gal inc=65 \& -110 & Fraction &  62\% & 28\% & 10\% \\
      \hline
      $r_{pericenter}$=24.8 & Stars & 2.72 & 2.31 & 0.78 \\
      mass ratio =3.0 & Gas & 3.34 & 0.12 & 0.11\\
      Gal inc=65 \& -89 & Fraction &  65\% & 26\% & 9\% \\
      \hline
      $r_{pericenter}$=24.8 & Stars & 2.38 & 2.67 & 0.95 \\
      mass ratio =2.8 & Gas & 2.95 & 0.13 & 0.10\\
      Gal inc=65 \& -89 & Fraction &  60\% & 29\% & 9\% \\
      \hline
      $r_{pericenter}$=24 & Stars & 2.25 & 2.84 & 0.88 \\
      mass ratio =2.5 & Gas & 2.91 & 0.12 & 0.09\\
      Gal inc=65 \& -89 & Fraction &  57\% & 32\% & 10\% \\
      \hline
      Same as above & Stars & 3.62 & 2.61 & 1.22 \\
      feedback & Gas & 1.32 & 0.07 & 0.12\\
      changed at 3.5Gyr & Fraction &  55\% & 29\% & 14\% \\
      \tableline
    \end{tabular}
    %% Any table notes must follow the \end{tabular} command.
  \end{center}
\end{table}

\subsection{Thin and thick disks: feedback prescriptions }

Table \ref{tbl-2} describes how the decomposition depends on the choice of orbital parameters. From the 2nd to the 4th sets of parameters, only the mass ratio varies and the bulge fraction increases as the mass ratio decreases as shown by \citet{hopkins09}. However, we have also tested a higher mass ratio (1st set of parameters) with a different set of inclinations for the progenitors. Thus the bulge fraction may also depend on other orbital parameters, here the relative inclination of the progenitors. For example, it reaches a maximum when the difference of inclination between the two progenitors is close to 180 degrees, possibly because more stars take radial orbits and fall into the bulge. It will be shown later in section 4.4 that this resonance has a more considerable effect on the matter that is stripped from the satellite to the tidal tail at the 2nd passage.

Almost all the simulations we have generated are re-forming significant thin disks containing more than half of the baryonic mass.  This is due to our feedback prescriptions that always preserve a significant fraction of gas just before the fusion.  As stated by \citet{cox08}, very little is known about the requisite conditions enabling the star formation to occur. An important prerequisite, however, is the presence of stable gas rich galactic disks at high redshifts, that could be the progenitors of M31. \citet{cox08} have tested feedback for an isolated disk galaxy with an initial gas fraction of $f_{gas}$=0.52. They found that the gas fraction is mostly preserved within one Gyr if median to high feedback prescriptions  are adopted, while low feedback decreases the gas fraction to  0.37 in the same amount of time.

In our current models we have assumed $f_{gas}$ $\sim$ 0.65: the re-formation of a significant thin disk implies that most of the gas is not transformed into stars before fusion. This calls for relatively high feedback, at least intermediate between the median and high values of \citet{cox08} We have verified that for half the value of high feedback of \citet{cox08}, most of the initial gas is preserved before the fusion. However, re-forming a significant stellar thin disk requires a significant change of the feedback at or after the fusion, in order to transform stars from the gas in the newly reformed disk. It results that from the merger of realistically gas-rich galaxies at z$\sim$ 1.4, we may reform a galaxy resembling M31, with the condition of a transitory history of the feedback during the merger, from high to low values. There could be some theoretical grounds favouring such a history, however. For example, in a gaseous rich and almost pristine medium, first supernovae are likely massive and generate high feedback as they may delay star formation for up to 100 Myr \citep[e.g.,][]{bromm09}. The fusion of two gaseous-rich galaxies corresponds to a severe mixing of most of their components, accompanied by large star formation rates. Then the metal abundance in the remnant becomes larger and more homogeneously distributed than it was in the progenitors, before fusion. The occurrence of very massive, primordial supernova is unlikely in such a mixed medium leading to a possible transition from high to low feedback. Before the fusion there are many fewer exchanges between the two interlopers, and it might be realistic to assume a negligible change in the feedback history.

Of course, the above is mostly made of conjectures in the absence of observations of gas-rich galaxies at high redshift showing no or very small amounts of star formation, i.e., the high-redshift counterparts of the present-day low surface brightness galaxies.  We may, however, verify whether our adopted star formation history is consistent or not with the star formation history revealed in each M31 substructure displayed in Fig. \ref{fig1}. The thick disk is indeed mostly made by matter returning from the tidal tails. It is comprised of approximately 80\% of stars older than 8 Gyr, the age of other stars being mostly from 5 to 8 Gyr. Such a distribution is well reproduced by our model with high feedback values (86\% of old stars) or with half this value (80\%). Because the thick disk shows a similar fraction of stars in the two bins set by \citet{brown08} at 5.5 and 7 Gyr, it is probable that there are no significant changes of the star formation history between the two corresponding epochs, i.e., between the first and second passages. 

\begin{figure}[!h]
  \centering
  \includegraphics [width=8cm] {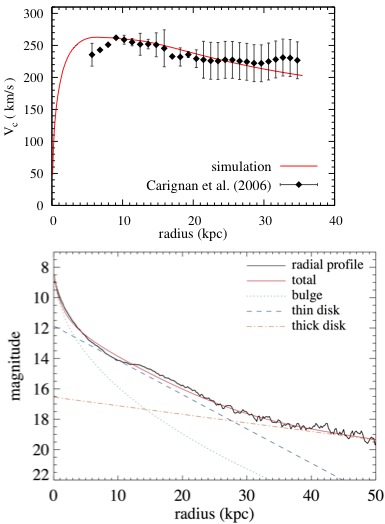} 
  \caption{3:1 merger with $r_{pericenter}$=24.8 kpc, 9 Gyr after the beginning of the simulation. {\it Top:} the rotation curve of the thin disk compared to \citet{carignan06}. There is a particularly good agreement between our modelling and data from \citet{carignan06}.
    %The curve in green has been re-scaled to account for the 8\% of baryonic matter which are not belonging to the system within 100 kpc. 
    {\it Bottom:} the decomposition of the mass profile in 3 components which evidences the 10kpc ring  and provides a thin disk scalelength of 5 kpc.}
  \label{fig5}
\end{figure}

Within these prescriptions our models predict quite well the fraction of each sub-component of  M31 (see Table \ref{tbl-2}). Note also that the modelled thin disks show rotation curves and scale lengths that are in good agreement with observations (see Fig. \ref{fig5}). In the following sections we examine whether this model could also reproduce other large-scale structures such as the ring and the Giant Stream.   

\begin{figure}[!h]
  \centering
  \includegraphics[width=8cm] {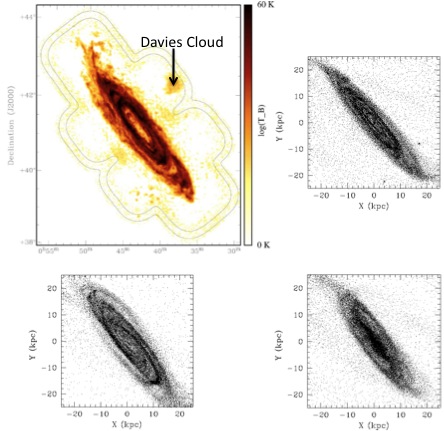}
  \caption{Comparison of the HI gas from \citet{braun09} with modelling of gas particles for 3:1 mergers at T=9 Gyr. On the top-right and bottom-left models are with $r_{per}$=24.8kpc, and inc of 65 and -95 degrees for the main galaxy and the satellite. The two models have a feedback varying from high to low values near the fusion time, and they do not show gas excess in the central bulge. They only differ by a small variation of 5 degrees in their inclinations along the z-axis. On the bottom-left is shown a model with a constant high feedback and with a difference between the two galaxies inclination of 170 degrees. It illustrates that changing feedback is necessary to explain the absence of gas at the centre and that inclinations should not be too close to 180 degrees to avoid too large distortions of the final gaseous disk.}
  \label{fig6}
\end{figure}

\subsection{The 10 kpc ring}
The decomposition of the light profile was done by fitting the radial profile
  with a 3 components model (bulge, thin, thick disks). This provides a thin disk with a Sersic index of 1 and a scale-length varying from  4.8 to 5.1 kpc for models with mass ratio ranging from 3 to 2.8. Such values are slightly smaller than the 5.8$\pm$0.4 kpc that was adopted by \citet{hammer07} for the M31 disk.  All these decompositions (see Fig. \ref{fig5}) reveal the presence of a  prominent ring at 10 kpc which corresponds well to the observed ring.  Similarly all models reproduce the 10 kpc gaseous ring (see Fig. \ref{fig6}) as seen in HI observations. This is not unexpected as all the orbits are close to polar which favours the formation of such prominent and persistent structures. In fact, HI observations do not provide many constraints on our modelling. They only discard parameters for which the disk is too warped, i.e., implying that the difference between the two inclination angles of the progenitors should not be too close to 180 degrees. Moreover, we notice (see Fig. \ref{fig6}) that with a constant high feedback during the simulation, there is still gas within the bulge of the remnant galaxy.  However this gas is consumed in models that assume a transition to low feedback immediately after the fusion. 

%\begin{figure*}
%   \centering
%\includegraphics[width=15cm]{Fig7}
%\caption{Illustration of the Giant Stream formation using a model with 300K particles, for a 3:1 merger with $r_{per}$=24kpc, and inc of 70 and -100 degrees for the main galaxy and the satellite. The first snapshot shows the two galaxies just before the second passage (same colour symbols than in Fig. \ref{fig2}). The two galaxies have angle difference near the resonance (180 degrees) providing many particles stripped from the satellite at 4.2 Gyr (see the tidal tail in the bottom of the satellite), until the fusion at 4.5 Gyr. Some of these particles have velocity below the escape velocity of the remnant system and are gradually falling back to the galaxy. They are tracing loops (or tore) around the newly formed disk as indicated by the arrow in the bottom-left image. Loops are fed by newly particles falling back from the tidal tail and are persistent until 9.3 Gyr and later. The dotted rectangle in the 4.2 Gyr panel illustrates how we have selected the tidal tail particles. The resulting galaxy in the last panel is compared with the inserted M31 image at the same scale (from Ibata et al. 2005, an enlarged view of this insert is provided by the central panel of Fig. \ref{fig1}).}
%\label{fig7}
%\end{figure*}

\subsection{The Giant Stream and its kinematics}

The Giant Stream is a very faint stellar structure, which extends from 50 to 80-100 kpc from the M31 centre. We have modelled it as being caused by particles coming back from the tidal tail formed just before fusion. Fig. \ref{fig2} describes the formation of such a  structure that is aligned along the trajectory of the satellite which falls into the mass centre at the fusion, 4.5 Gyr after the beginning of the simulation. We verified that, within the family of orbits we choose, the strength of the tidal tail (see Fig. \ref{fig2}, panel at 4.2 Gyr) depends on the inclination of the progenitors relative to the orbital angular momentum (see Table \ref{tbl-1}). Optimal values are found for large values of the inclination of the main galaxy ($>$ 50 degrees) and especially for differences between the inclination of the two progenitors between 140 to 170 degrees, i.e., not too far from the resonance at 180 degree. 

\begin{figure}[!h]
  \centering
  \includegraphics[width=8cm]{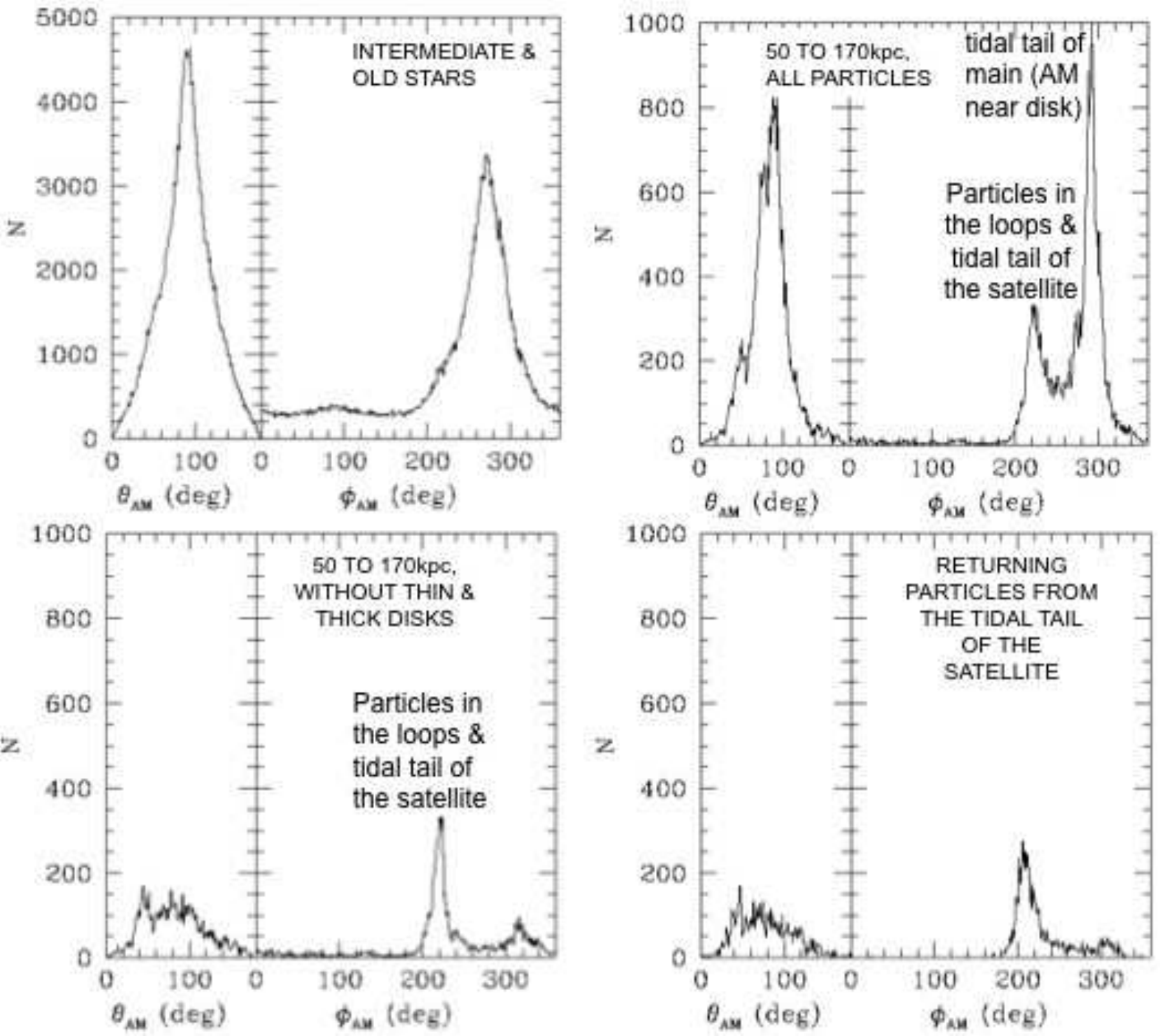}
  \caption{2.8:1 merger with $r_{pericenter}$=30 kpc. The distribution of the angular momentum 5 Gyr after fusion using spherical coordinates, assuming the thin disk perpendicular to the y-axis. {\it Top-left:} stars formed before fusion (intermediate and old stars) which show a broader distribution than thin disk stars as it is expected for the thick disk that hosts most of these stars (see Table \ref{tbl-2}). {\it Top-right:} angular momentum distribution for particles beyond 50 kpc from the centre of the remnant.  Most particles in the outskirts have angular momentum following the two tidal tails, one associated with the main progenitor, the other associated with the satellite. {\it Bottom-left:} same after removing both thin and thick disks: only the particles in the loops are kept.  {\it Bottom-right:} the distribution of angular momentum for all the particles associated with the tidal tail of the satellite, evidencing that these particles are lying in the  $\Phi$=225 degrees plane.  }
  \label{fig7}
\end{figure}

Modelling the Giant Stream is not an easy task as the structure is very faint compared to other large substructures of M31, even the thick disk. As in section 4.1 we make use of angular momentum properties to identify peculiar structures able to persist after the remnant phase (see Fig. \ref{fig7}). 

Fig. \ref{fig7} shows the distribution of the angular momentum  for the thick disk (top-left panel) and at the galaxy outskirts, 5 Gyr after fusion. The top-right panel shows two peaks which correspond to the two tidal tails formed after the second passage. The tidal tail associated to the main galaxy (see Fig. \ref{fig2} at 4.2 or at 4.5 Gyr)  is spiralling around the thick disk and its angular momentum peaks at $\Phi$=290 degrees, which is indeed part of the thick disk (see top-left panel). This peak disappears in the bottom-left panel for which the thin and thick disks have been removed. The residual particles form another well identified peak at $\Phi$=225 degrees. We have verified that they correspond to particles returning from the tidal tail of the satellite: in the bottom-right panel, selected  particles of the tidal tail of the satellite indeed show a very similar distribution with a peak at $\Phi$=225 degrees, which is offset by 45 degrees from the disk angular momentum. The selection of these particles has been done in a rather crude way (see Fig. \ref{fig2}, panel at T=4.2 Gyr): we simply "cut the tidal tail" at the time of its formation, up to the edge of the satellite. However this crude selection misses some particles, for example those which are stripped from the satellite at later phases (e.g. 3rd passage, see Fig. \ref{fig2}, panel at 4.5 Gyr). Comparing the counts of particles having an angular momentum around the  $\Phi$=225 degrees peak, we find that our selection recovers  90\%, 70\%, 40\% and 15\% within shells of 100-170kpc, 50-100kpc, 30-50kpc and 20-30kpc, respectively.  As expected, particles at the very edge of the tidal tail are the most difficult to pre-select. Five gigayears after fusion those particles have returned to trajectories close to the galaxy centre.

\begin{figure*}[!ht]
  \centering
  \includegraphics[width=15cm]{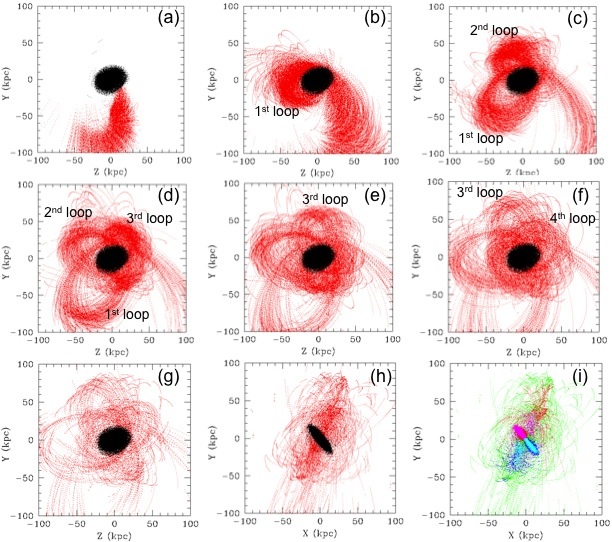}
  \caption{2.8:1 merger with $r_{pericenter}$=30 kpc. {\it From panels (a) to (g):} Snapshots in the (y, z) plane of trajectories for particles (red dots) which are part of the tidal tail of the satellite. Panel (a) shows the system just before fusion when particles are stripped from the satellite. Panels (b) to (f) show the particle motions for time intervals of 0-1Gyr, 1-2Gyr, 2-3Gyr, 3-4Gyr and 4-5Gyr after fusion, respectively. Panel (g) shows the distribution of particles 5.5 Gyr after fusion. For illustration, we have added  in all panels  the thin disk (black) as it is, 5.5Gyr after fusion. It shows the formation of the different loops which are drawn by particles when coming back from the tidal tail. {\it Panels (h) and (i):} Same as panel (g) but projected in the observer's frame in the (x, y) directions illustrating that the loops are inserted within a thick plane at 45 degree from the disk PA. Panel (h) shows the same particles as panel (i) for which the heliocentric velocities are coded following \citet{chemin09}, i.e. cyan ($<$-500 km/s), blue (-500 to -400 km/s), green (-400 to -200km/s), red (-200 to -100 km/s) and magenta ($>$ -100km/s).  It illustrates a disk rotation very similar to the observed one as well as the velocity of the Giant Stream which is reaching larger negative velocities when it reaches the M31 centre. }
  \label{fig8}
\end{figure*}

Tracing the trajectories of selected particles in the tidal tail is very instructive: they correspond to an important component of the angular momentum which is at odd with the rotation of both the thick and thin disks (see Fig. \ref{fig8}).

Fig. \ref{fig8} describes the temporal evolution of the trajectories of tidal tail particles returning to the galaxy. When falling to the galaxy centre, the particles are looping around the galaxy with a pericenter at $\sim$ 25 kpc. These particles are describing several elliptical orbits -or loops- around the galaxy potential. We identified 4 of these loops, although they might be more numerous. Fig. \ref{fig8} shows that with increasing time, they show a precession around the galaxy as well as they are growing. This is because stars coming back from the tidal tail are returning from higher elevation at a later epoch and are coming back with increasing energy.  We also note that the loops are somewhat thick and, as such, they could better be called tores.

The permanent rain of stars from the tidal tail ensures very long-term streams of stars within the galaxy outskirts well after the remnant phase. Infalling stars are trapped into elliptical orbits with the remnant galaxy right at the foci, which make the orbit stable. These orbits are those expected for extremely small satellites (single stars!), in the absence of tidally induced forces. Loops are expected to persists  for several billion years\footnote{We have verified that the resulting dark matter density profile in the remnant is slightly oblate (e=0.2) shortly after the fusion of the two galaxies and then stable: then it could not strongly affect the trajectories of returning star particles from the tidal tail.} after fusion and, moreover, they are permanently fed by new particles coming from the tidal tail. These streams and loops are all within a common plane, 45 degrees from the PA of the rejuvenated disk as it is evidenced by Fig. \ref{fig8} (panels h and i). This is our proposed model for the formation of the Giant Stream which indeed points toward the centre of the galaxy. Let us now examine how this model is compliant with several observations of the Giant Stream properties.

\begin{figure}[!h]
  \centering
  \includegraphics[width=8cm]{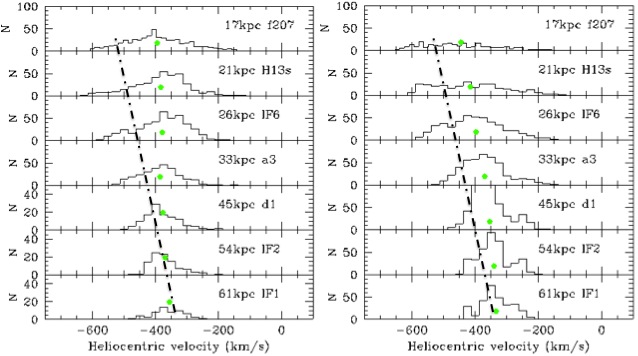}
  \caption{{\it Left:} same model as in Fig 3. The distribution of heliocentric velocities in the Giant Stream region at distances indicated in each panels with references to fields from \citet{gilbert09}. The dot-dashed line reproduces the  black line of Figure \ref{fig1} of \citet{ibata04} representing the  high negative velocity edge. Green dots represent the mass weighted average for each panel. {\it Right:} same for particles returning from the tidal tail.  In this plot we have accounted for several snapshots in the simulation to artificially increase the number of stars in the central regions (see the text).}
  \label{fig9}
\end{figure}

\begin{figure}[!h]
  \centering
  \includegraphics[width=8cm]{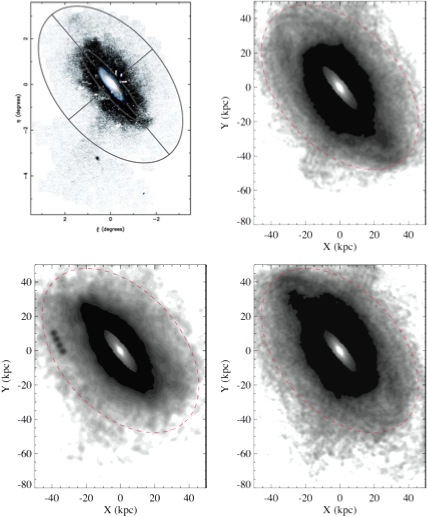}
  \caption{Comparison of observations \citep[top-left from][]{ibata05} with our modelling; at top-right, we adopted a mass ratio of 2.8:1 with $r_{per}$=30 kpc observed after 10 Gyr, at bottom-left, a mass ratio of 3:1 with $r_{per}$=25 kpc observed after 9 Gyr. The bottom-right panel shows the same as that of the top-right, except that we have inverted all axes. 4 to 10 snapshots have been used to optimise the contrast.}
  \label{fig10}
\end{figure}

The kinematics of stars follows the trend evidenced by \citet{ibata04} and \citet{gilbert09}: stellar particles in the streams reach large negative values of their heliocentric velocities towards the centre (see panel i of Fig. \ref{fig8}). Fig. \ref{fig9} shows the distribution of heliocentric velocities for stellar particles selected in the region of the Giant Stream within $\Phi_{obs}$ from -90 to -140 degrees. Observations (see Fig. \ref{fig1} of \citet{ibata04} and Fig. \ref{fig5} of \citet{gilbert09}) show a broadening of the velocity distribution towards the galaxy centre that is well reproduced by our simulations. Besides this, our model also well reproduces the distribution of stellar ages observed by \citet{brown08}, as we find 80\%, 20\% and 0\% of stars older than 8 Gyr, from 5 to 8 Gyr and younger than 5 Gyr, respectively. These fractions can be regulated by changing the feedback prescriptions (see section 4.2).

We have attempted to reproduce the photometric properties of the Giant Stream. The difficulty of such an attempt is due to the fact that the Giant Stream is very faint when compared to other sub-structures of M31, which is a problem also encountered by observers. Even with 500 000 particles, the number of stellar particles in the Giant Stream  is below 1000 which obviously limits a detailed reproduction of the structure. Within an angle of $\Phi_{obs}$ from -90 to -140 degrees and from 25 to 80 kpc, the total mass of the Giant Stream is 4.8 and 4.4 $10^{8}$ $M_{\odot}$ in the two examples illustrated in Fig. \ref{fig10}. Using other variations of the model parameters we find a range from 2 to 5$\times$$10^{8}$ $M_{\odot}$, which easily includes the observational value of 3 $10^{8}$ $M_{\odot}$ \citep[see][]{mori08}.

All our models (column 4 of Table \ref{tbl-1}) show streams of stars which belong to a plane that intersects the observational plane in a direction 45 degrees from the thin disk PA. We verify that to optimise the reproduction of the observed features in position and strength, we can use two different kinds of symmetries. First, we can rotate the whole system around an axis perpendicular to the thin disk, as this would not affect the galaxy decomposition in substructures, nor the kinematics of thin and thick disks. Second, we find that gravity is not sensitive to a complete inversion of the initial (x, y, z) coordinate system. This has the advantage of being able to reproduce the Giant Stream with many variations of our model parameters, but it also has the inconvenience of providing a very large amount of space parameters to investigate. Fig. \ref{fig10} illustrates some examples of this exercise for two models with 500K particles for which the Giant Stream mass is slightly larger than the observed one. Much larger number of particles are certainly required to reproduce at the same time all the morphological details of the thick disk, although we generally reproduce quite well the NE and W shelfs.

\section{Discussion and Conclusion}

By construction, all models reproduce quite well the distribution of stellar ages in the various substructures, as they are described in Fig. \ref{fig1}. This is especially true for both the thick disk and the Giant Stream (see sections 4.2 and 4.4) for which our predicted compositions of stellar ages are from 10\% to 20\% of intermediate age stars (5-8 Gyr), the rest essentially being made of older stars. However, we experienced some difficulties to reproduce the distribution of the 21kpc and 35kpc "Halo" fields of \citet{brown08}. We indeed find a fraction of intermediate age stars ranging from 5\% to 15\% , while according to  Figure \ref{fig1}, it should be near or below  5\%. It would be tempting to use this property for selecting the best model of M31. However, in our models intermediate age stars are related to low mass particles, three of them being produced by a single gas particle. It may lead to an artificially scattering of those intermediate age star particles (see section 3), which could be only solved by simulations with a much larger number of particles than 500K.

All models as defined in Table \ref{tbl-1}, column 4 can reproduce with a quite good accuracy the bulge, and thin and thick disk fractions, as well as the ring and the shape, the mass, and the kinematics of the Giant Stream. The examples shown in Fig. \ref{fig10} are cases for which the Giant Stream stellar content is slightly larger than the observed one, for illustration purposes. We also note that the shapes of the predicted thick disk and Giant Stream can significantly change -within a given set of orbital parameters (see Table \ref{tbl-1})- either because they can be modified through a rotation around the disk axis or because they could fluctuate with the assumed time after the beginning of the simulation. It is beyond the scope of this paper to search for the best model that reproduces all the details of M31 substructures. This would require launching several hundreds of simulations with several millions of particles. We have, however, learned from this analysis that:

\begin{itemize}
\item A thin disk such as that of M31  can be reproduced with a combination of $\sim$ 65\% of gas in the progenitors and a star formation history that prevents gas consumption before the fusion; low efficiency of star formation is thus predicted in primordial gaseous phases, while when the medium has been sufficiently enriched and mixed during fusion, the gas is more easily consumed; this transition has essentially been modulated through the assumed feedback history; 
\item The same star formation history also reproduces a rotationally-supported thick disk with a stellar composition similar to what is observed; and
\item also fractions of bulge/thin disk and thick disk similar to what is observed within 10\% accuracy.
\item These polar orbits always reproduce the observed 10kpc ring, and in most cases, the gas map is very similar to the observed one (see Fig. \ref{fig6}), especially when the gas in the galaxy centre is allowed to form stars after the galaxy fusion;
\item The predicted Giant Stream is made of stellar streams due to returning stars from a tidal tail formed prior to fusion from stripped particles of the satellite; these stars are drawing streams along loops which are inserted into a thick plane that is 45 degrees inclined relative to the disk PA (see Fig. \ref{fig8}). Such streams are very long-lived because returning particles are trapped into such loops for periods larger than several gigayears, and also because the streams are permanently fed by new particles falling back from the tidal tail at all epochs.
\end{itemize}

The success of modeling M31 properties as resulting from a past major merger event is patent. Its predictive power is however limited by the complexity of the system as well as the very demanding number of particles required to reproduce in details all substructures of M31. An important limitation could also be the accuracy of the calculations, which requires tracking with an unprecedented accuracy positions and velocities of particles for almost a Hubble time. In fact, we cannot claim to have determined the best major merger model of M31, although it is likely within the limits discussed below.

We did not either try to investigate the whole parameter space to which the above results hold, although examination of Table \ref{tbl-1} led us to suspect that it is likely large. However some specific links between parameters and observations may provide some empirical limits. The bulge fraction can be reached within a 10\% accuracy for all models  with the mass ratio of 3$\pm$0.5. In the frame of our model for M31, the halo is enriched by stars coming from the progenitors before the interaction, while the thick disk and Giant Stream are further enriched by stars formed before the first passage and fusion. This (see Figure \ref{fig1} and section 2) is compliant with pericenter radii of 25$\pm$5 kpc. Further modelling and measurements of stellar ages in several fields in the outskirts of M31 would certainly help to provide more accurate values. This would also provide better accuracy for the important epochs of the interaction which would have begun 8.75$\pm$0.75 Gyr ago with a fusion time 5.5$\pm$0.5 Gyr ago. There is still work needed to properly constraint other orbital parameters such as the inclinations of progenitors relative to the angular momentum direction. We are far from having investigated the whole range of orbital properties which produces a strong tidal tail from stripped material from the satellite, and then a structure similar to the Giant Stream. Table \ref{tbl-3} describes all the models presented in this paper, which are part of about 50 tested models. Fig. \ref{fig6} demonstrates that models with differences of progenitor inclinations near 180 degrees show not only a maximum of resonnance (thus a very strong tidal tail) but also galactic disks which are probably too warped or distorted. We have mostly investigated models with the main galaxy inclination of 55 to 70 degrees, but there could be other solutions for different angles. All the models in Table \ref{tbl-3}, as well as models in the range described by Table \ref{tbl-1} (4th column) reproduce quite well the observations, including the Giant Stream.  A much more detailed analysis is necessary to determine the best model parameters, and this could be done by comparing in detail the constraints provided by the gaseous disk, the thick disk, and the Giant Stream shapes as well as by any other structures discovered in the M31 halo field.

\begin{table*}[!ht]
  \begin{center}
    \caption{Summary of model parameters from Table \ref{tbl-1} used throughout the paper (all within the range of parameters of Table \ref{tbl-1}, column 4). Angle, pericenter radii, number and mass of particles are  given in degrees, kpc, 1000s of particles and  $(10^6 M_{\odot})$, respectively. Feedback adopted values are (1): high feedback; (2): 5 times medium feedback; (3) high feedback before fusion then low feedback and (4): 5 times medium feedback before fusion then low feedback. Model parameters for Fig. \ref{fig4}, \ref{fig7}, \ref{fig8}, \ref{fig9} and \ref{fig10} are the same than for Fig. \ref{fig3}; the other model used in Fig. \ref{fig10} is the same as that of Fig. \ref{fig11}a. The 6 last lines of the Table provides a check of the goodness of each model within $\pm$20\% of the observed value: the decomposition of M31 in the bulge, and the thin and thick disks (e.g. B/T=0.28$\pm$0.05), the disk scale length (e.g. 5.8$\pm$1.1 kpc), the 10 kpc ring, the Giant Stream (e.g. 3$\pm$0.6 $10^{8}$ $M_{\odot}$), the stellar ages in the thick disk and in the Giant Stream (assuming fraction of $>$ 8 Gyr stars to be 85$\pm$8\%) and gas fraction in the galaxy within 30 kpc (7$\pm$2\%). }
    
    {\scriptsize
      \begin{tabular}{llllllllllll}
        \tableline\tableline
        parameter         & Fig2 & Fig3     & Tab2a     & Tab2b &Tab2c     & Tab2d     &Tab2e      & Fig5      &Fig6a&Fig6c      & Fig11b\\
        &  &      &      &  &     &      &      & \& Fig6b & \& Fig11a &        & \\
        \tableline                                                                                                                                              
        mass ratio        &   3.0   & 2.8      & 3.5       & 3.0  & 2.8       &  2.5      &2.5        &3.0        & 3.0   &3.0     & 2.8   \\
        %$f_{gas}$ Gal1    &      & 0.6      & 0.6       & 0.6  & 0.6       &  0.6      &0.6        &0.6        & 0.6 &0.6  & 0.6 & 0.6   & 0.6        & 0.6   \\
        %$f_{gas}$ Gal2    &      & 0.8      & 0.8       & 0.8  & 0.8       &  0.8      &0.8        &0.8        & 0.8 &0.8  &0.8  & 0.8   & 0.8        & 0.8   \\
        Gal1 incy         &  70    & 65       & 60        & 65   & 65        &  65       &65         &65         & 65    & 65           & 65    \\
        Gal2 incy         &  -100    & -89      &-109       & -89  & -89       & -89       &-89        &-95        &-95   &-105          & -89   \\
        Gal1 incz         &  50    & 90       & 110       & 90   & 90        &  90       &90         &90         & 85    & 90      & 90    \\
        Gal2 incz         &   90   & 90       & 110       & 90   & 90        & 90        &90         &90         & 85     & 90     & 90    \\
        $r_{pericenter}$  &  24    & 30       & 24        &24.8  & 24.8      & 24        &24         &24.8       & 24.8 & 26     & 30    \\
        Feedback          &  1    & 1     & 1      & 1 & 1      & 1      & 3   & 3   & 4 & 1  & 1  \\
        %$c_{star}$        &      & 0.03     & 0.03      &0.03  & 0.03      & 0.03      &0.03       &0.03       &0.03 &0.03 &0.03 & 0.03  &0.03        & 0.03  \\
        N$_{particle}$    &  300    & 300     & 154      &300 & 159      & 300      &159      &540       &540  &300    & 960  \\
        %N$_{newstar}$     &    3  & 3        & 3         & 3    & 3         & 3         &3          &2          & 2   & 2   &  3  & 2     & 2          & 3      \\
        %$M_d:M_s:M_g:M_{ns}$&    &2:1:1:0.33&4:1:0.5:0.17&2:1:1:0.33&4:1:0.5:0.17&2:1:1:0.33&2:1:1:0.33 &2:1:1:0.5  &2:1:1:0.5&2:1:1:0.5&     & 4:1:1:0.5&2:1:1:0.5&2:1:1:0.33\\
        M$_{DM}$ &  2.2 & 2.2  & 7.3       & 2.2  & 7.3       & 2.2       &7.3        &1.2        & 1.2  & 2.2 & 0.68   \\
        M$_{old star}$ &  1.1  & 1.1      & 1.8       & 1.1  & 1.8       & 1.1       &1.8        &0.6        & 0.6 & 1.1   & 0.34   \\
        M$_{gas}$  &   1.1   & 1.1      & 0.9       & 1.1  & 0.9       & 1.1       &0.9        &0.6        & 0.6  & 1.1   & 0.34   \\
        M$_{new star}$  &  0.37    & 0.37     & 0.3       & 0.37 & 0.3       & 0.37      &0.3        &0.3        & 0.3  & 0.37     & 0.11   \\
        \tableline
        model goodness &  &      &      &  &     &      &      &       &  &        & \\
        decomposition &  OK &   OK   &  OK    & OK  &  OK   &   OK   &  NO    &   OK    & OK & OK    & OK\\
        disk scale length &  OK &   OK   &  OK    & OK  &  OK   &   OK   &  OK    &   OK    & OK & OK    & OK\\
        10kpc ring  &  OK &   OK   &  OK    & OK  &  OK   &   OK   &  OK    &   OK    & OK & OK  & OK\\
        Giant Stream &  OK &   OK   &  OK    & OK  &  OK   &   OK   &  OK    &   OK    & OK & OK    & OK\\
        stellar ages &  OK &   OK   &  OK    & OK  &  OK   &   OK   &  OK    &   OK    & OK & OK    & OK\\
        gas fraction &  NO\tablenotemark{a}  &   NO\tablenotemark{a}    &  NO\tablenotemark{a}    & NO\tablenotemark{a}   &  NO\tablenotemark{a}    &   NO\tablenotemark{a}    &  OK    &   OK    & OK & NO\tablenotemark{a}  & NO\tablenotemark{a} \\
        \tableline
      \end{tabular}}
    \label{tbl-3}
    \tablenotetext{a}{Models with a constant star formation with low efficiency (or high feedback) could not transform enough gas into stars after the fusion leading to a more gas-rich disk than observed; assuming a varying star formation (more efficient after the fusion) suffices to correct the discrepancy.}
    
  \end{center}
\end{table*}

\begin{figure}[!h]
  \centering
  \includegraphics[width=8cm]{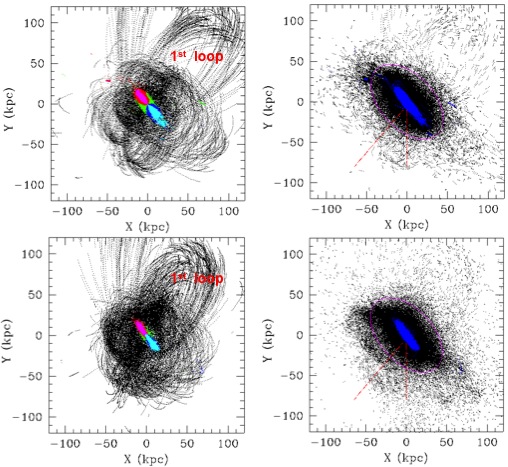}
  \caption{Two models for which most features have been reproduced including those from the new PANDA survey. {\it Top}: a model with 500K particles assuming a 3:1 merger, $r_{per}$=25 kpc, and inversion of the initial (x, y, z) coordinate system. On the left are shown the returning particles from the tidal tail as in Fig. \ref{fig8}, panels (h) and (i), with black dots and with the rotating disk inserted (same code as that of Fig. \ref{fig8}). On the right all particles of the simulation are shown, including 4 snapshots around T=8.72 Gyr. This model is with $Gal1_{incy}$=65, $Gal2_{incy}$=-95, $Gal1_{incz}$=85, $Gal2_{incz}$=85, half the maximal feedback until 0.5Gyr after fusion, then low feedback. {\it Bottom}: a model with 1M particles assuming a 2.8:1 merger, $r_{per}$=30 kpc and inversion of the initial (x, y, z) coordinate system. Here only 2 snapshots around T=9.05Gyr are shown in the left image. Here a constant high feedback history has been assumed and other parameters are $Gal1_{incy}$=65, $Gal2_{incy}$=-89, $Gal1_{incz}$=110, $Gal2_{incz}$=110. The Giant Stream mass includes 2.5 to 2.8 $10^{8}$ $M_{\odot}$.}
  \label{fig11}
\end{figure}

While writing this paper, we become aware of the discovery of many other structures in the halo of M31 \citep{mackey10}, especially after the successful deep imagery of the whole field surrounding M31 by the PANDA team \citep{mcconnachie10}. This magnificent image reveals the presence of loops surrounding M31 which strikingly resemble our model predictions (see Fig. \ref{fig8}). In particular the northern loop found by the PANDA team which extents up to 120 kpc from the galaxy centre provides particularly strong constraints on our model. Indeed the first loop in Fig. \ref{fig8} can accurately reproduce the northern loop \citep[called NW stream in the incomplete image of][]{mackey10}, and other loops are well matched with the Eastern arc and Stream  C. Fig. \ref{fig11} shows two examples of models that succeed in reproducing most of these features along with the Giant Stream and most other substructures (thin disk, bulge, and thick disk) of M31. In fact we also verified that the Giant Stream kinematics  is still reproduced. Fitting these additional structures allows us to fix the rotation angle around the thin disk axis as well as solving the initial conditions of the system because only the inversion of the initial (x, y, z) coordinate system is able to reproduce the ensemble of features in Fig. \ref{fig11}. The size of the loop also depends on the time elapsed after fusion and the beginning of the interaction is expected to be 8.75$\pm$0.35 Gyr within the range of parameters in Table \ref{tbl-1}.

As a whole, M31 and the complex structures in its halo can be reproduced by assuming a single major merger which may have begun 8.7 Gyr ago. The advantages of such a solution for M31 compared to the hypothesis of numerous minor mergers are as follows:
\begin{itemize}
\item It explains most of the complexities of M31 by a single event rather than numerous and minor merger events which have to be adapted in a somewhat ad-hoc way to each feature discovered in the M31 halo, including the thick disk;
\item It naturally explains  stellar ages and metallicities of most substructures in the M31 halo as well as why they show so many similarities;
\item It overcomes the increasing difficulty of identifying the residuals of satellites assumed to be responsible for the different features in the M31 halo; 
\item It is consistent with the kinematics of the M31 globular system \citep[e.g.,][]{bekki10}.
\end{itemize}

\begin{figure}[!h]
  \centering
  \includegraphics[width=7cm]{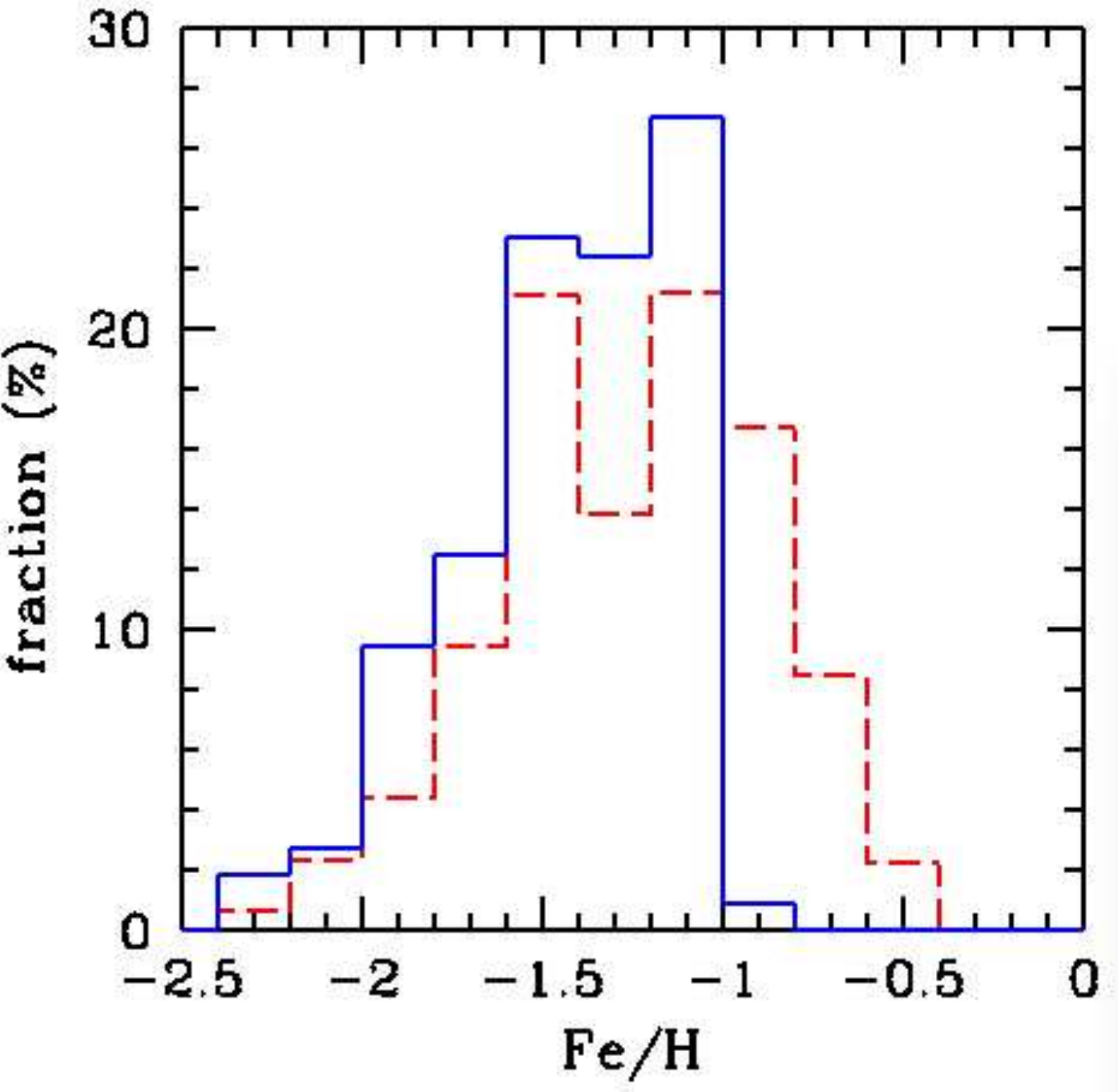}
  \caption{Fe/H distribution of progenitor stars in the Giant Stream (red dashed line) and in the northern loop (blue solid line), for the same model as that of Fig. \ref{fig11}a. The absence of metal rich stars (e.g., -0.4$<$Fe/H$<$-0) is due to the fact that particles returning from the tidal tail are coming from the outskirts of the progenitors. The absence of stars richer than Fe/H=-0.7 could explain why the northern loop is not detected with such a metallicity selection, while it becomes prominent for metal poor stars (Fe/H$<$-1.4).}
  \label{fig12}
\end{figure}

Re-formation of disks is expected after gas-rich major mergers \citep{barnes02, springel05b}. It is more than just a curiosity since the disk rebuilding scenario has been proposed as a channel for the general formation of spirals \citep{hammer05}, including specifically the case of M31 \citep{hammer07}. Gas-richness of progenitors is certainly high, as it has been estimated for many high-z galaxies. The requirement of a star formation with a low efficiency before the fusion can be reached through any mechanism that prevents star formation in relatively primordial medium. It could be caused either by higher feedback efficiency in primordial medium as assumed in section 4.2, or because the gas in the progenitors is less concentrated than in present-day spirals as it is in present-day low surface brightness galaxies, or because cooling is less efficient in a relatively pristine medium, or a combination of all these factors. More generally, the expected increase of the gas metal abundance (expected to be slow before the fusion but very efficient during the fusion) may help to increase the molecular gas fraction, the optical depth of the gas and the radiation pressure effects, all contributing to a change in the star formation history during the interaction (TJ Cox, private communication). The above is quite well illustrated {\it a contrario} by \citet{bournaud10} who assumed that progenitors of high-z major mergers are very strong star forming, turbulent, gas-rich disks: they naturally find that remnants cannot be disk dominated within such conditions.  

%The rebuilding disk scenario as well as our proposition for M31 past history are probably at odds with the presence of highly star-forming disks at high redshift (Genzel et al. 2006; see also Forster Schreiber et al. 2009). However it has not been solved yet whether or not these high-z objects would still be disks after using the simple criterion\footnote{at low spatial resolution, a shear in the velocity field if due to rotation should show a peak in the dispersion map between the velocity extrema} established by Flores et al. (2006) and indeed some of them might be simply major mergers (Robertson et al. 2008). 

There is still a considerable need to improve the modelling of M31, by using a much larger number of particles as well as by providing more kinematical and stellar age constraints in the numerous fields of view surrounding M31. We also need to reproduce with more accuracy the shape of the Giant Stream and of the thick disk contours. Metal abundance enrichment can  also be implemented to more accurately fit the observations. For example, the striking differences between features at different metallicity in the halo found by the PANDA team cannot be retrieved by our modelling in the absence of such an implementation\footnote{This may also help to better reproduce the 35kpc halo field of \citet{brown07} which is in-between the Eastern arc and the stream C \citep{mackey10}: this also points out the need for resolving stellar streams and then for a better accuracy in following positions and velocities of particles; a similar need is for identifying carefully the particles in loops such as that proposed for the Northern loop. }.
We have attempted to verify whether such a trend is predicted by our model. At first order, one may "paint" each star in the progenitors assuming a reasonable metallicity gradient . We adopt the approach by \citet{rocha08} assuming that metallicity gradient follows the dust gradient, with a scalelength 1.4 times larger than that of the stars. The range of metal abundance (Fe/H) is adopted from 0 to -2.5. The stellar metal abundance of the satellite is likely significantly smaller than that of the main progenitor, because their stellar mass ratio (for our 3:1 models, see Table \ref{tbl-1}) is 5. Little is known about the shape of the metal-stellar mass relationship at z $\sim$ 1.5, although a serious attempt has been made at z$\sim$ 0.7 by \citet{Rodrigues08}. Assuming this holds up to z=1.5, stars in the main progenitor are on average 0.4 dex more metal rich than those from the satellite, a value smaller than the large scatter of the \citet{Rodrigues08} relationship. Figure \ref{fig12} shows the metallicity distribution of "painted" stars in both the Giant Stream (defined as in section 4.4, from 25 to 80 kpc and $\Phi_{obs}$ from -90 to -140 degrees) and  in the northern loop (defined from 55 to 130 kpc and $\Phi_{obs}$ from 0 to 90 degrees) discovered by the PANDA team. Because most of the stars in the northern loop are coming from the satellite, this could explain why this feature is not detected in the metal-rich map of the PANDA team (see Fig. \ref{fig12}). Such a prediction may stand unless intermediate age stars (those formed before the fusion) strongly affect the metal abundance of both the Giant Stream and the northern loop. On the other hand the same model correctly predicts the (observed) small fraction of such stars in the Giant Stream. 

%On the other hand we are quite optimistic about the possible outcome. The particles in the Northern loop should have been the latest stars returning from the loop and as such, are likely stars stripped from the outskirts of the satellite just before fusion: it is natural to expect that such stars are coming from less enriched material than stars stripped from the satellite central regions, those ones populating the Giant Stream and the Eastern arc. 

Another important improvement in this modelling would be to consider the possible interaction of M33 which has been proposed to be responsible for several features in the SE of the M31 halo by \citet{mcconnachie09}. Although the dark matter density profile does not significantly affect our results (results shown in Fig. \ref{fig2}, \ref{fig3}, \ref{fig4}, \ref{fig5}, \ref{fig7}, \ref{fig8}, \ref{fig9}, and \ref{fig11} are similar when using a Hernquist profile instead of a core profile), we also need to verify whether these are affected by adopting different baryonic mass fractions, especially for  lower values than our adopted 20\%.  Indeed, a considerable change of the dark matter fraction may prevent stars from being stripped by the collision and then the distribution of baryonic matter in the outskirts may be significantly affected.

However, we believe the result of such a model is very encouraging because it fulfills the Occam's razor principle: a single event may explain most -if not all- the properties of M31 and of its outskirts. It also proposes a mechanism for the Giant Stream which is simply fed by stars which are captured into loops orbiting around M31 for elapsed times which may reach a Hubble time. If M31 is actually the result of an ancient major merger, there will be a considerable need to revise our knowledge about our immediate environment and in cosmology\footnote{The history of galaxy formation theory has been strongly influenced by our knowledge of the Milky Way and the theory of spiral galaxy formation a la \citet{ELS62} is still lively; the fact that the second closest galaxy, M31, could plausibly be a major merger may remind us that the theory of galaxy formation could require significant adjustments.}, for the following reasons:

\begin{itemize}
\item  Merging of gas-rich distant galaxies can easily produce large thin disks by assuming a less efficient star formation before the fusion; this simple scheme supports the rebuilding disk scenario for many giant spiral galaxies at least for those sharing similar properties (enriched halo stars) as M31;
\item Most models show the formation of tidal dwarves, including those from tidal tails formed after the first passage, and some of them could be part of the satellite system of M31;
their properties, after comparison with observations,  may be used in the future as a constraint to the orbital parameters in further modelling;
\item Up to 15\% of the material is ejected after the fusion and this may populate the whole Local Group including in the direction of the Milky Way; this leads \citet{yang10} to investigate whether the Large Magellanic Cloud could be related to some ejected material coming from M31.
%\item there could be a need for a radical revision of the M31 and of the Milky Way total masses as those have been often estimated assuming that their satellite system is relaxed; thus the total mass of the Local group may be revised downwards and be more compliant with cosmological ratio of baryonic to total mass in an environment which could be typical of a median galaxy density in the field. 
\end{itemize}

{\it Note added into proofs:} The specific frequency of globular clusters in M31 appears
to be three or four times greater than it is in the Galaxy \cite{vandenBergh10}, i.e., it is expected if a major
merger occurred in the first galaxy.  The fact that the globulars in M31 and in the Galaxy have almost exactly the
same half light radii  supports the
notion that M31 formed by merger of two Galaxy-sized ancestors (Sidney van den Bergh, private communication) .

%\section{Discussion}

%\section{Conclusion}

%\begin{figure}
 % \centering
%\includegraphics[width=9cm]{GS_location.jpg}
%\caption{The Giant Stream and its radio counterpart.}
%\label{fig1}
%\end{figure}

\acknowledgments
This work has been supported by the China-France International Associated Laboratory "Origins" and by the National Basic Research Program of China (973 Program), No.~2010CB833000. Part of the simulations have been carried out at the High Performance Computing Center at National Astronomical Observatories, Chinese Academy of Sciences as well as at the Computing Center at Paris Observatory. We warmly thank Tom Brown, TJ Cox and Sidney van den Bergh for their very useful comments during the submission process of this paper. Suggestions and comments from an unknown referee have been very helpful in improving the current version of the article.

\clearpage

%% the file, uncomment the \plotone, \plottwo, and
%% If you need a layout that cannot be achieved with \plotone or
%% \plottwo, you can invoke the graphicx package directly with the
%% \includegraphics command or use \plotfiddle. For more information,
%% please see the tutorial on "Using Electronic Art with AASTeX" in the
%% documentation section at the AASTeX Web site,
%% http://www.journals.uchicago.edu/AAS/AASTeX.
%%
%% The examples below also include sample markup for submission of
%% supplemental electronic materials. As always, be sure to check
%% the instructions to authors for the journal you are submitting to
%% for specific submissions guidelines as they vary from

%% This example uses \plotone to include an EPS file scaled to
%% 80% of its natural size with \epsscale. Its caption
%% has been written to indicate that additional figure parts will be
%% available in the electronic journal.

\clearpage

%% Here we use \plottwo to present two versions of the same figure,
%% one in black and white for print the other in RGB color
%% for online presentation. Note that the caption indicates
%% that a color version of the figure will be available online.
%%

\end{document}